\documentclass{svjour3}                     
\smartqed  
\usepackage{graphicx}
\journalname{Few Body Systems}
\begin{document}

\title{A relativistic coupled-channel formalism for electromagnetic form factors of 2-body bound states
\thanks{Presented at the 21st European Conference on Few-Body
Problems in Physics, Salamanca, Spain, 30 August - 3 September
2010.} }


\author{
E. P. Biernat\and W. H. Klink\and W. Schweiger 
}


\institute{E. P. Biernat  \and W. Schweiger \at
              Institut f\"ur Physik, FB Theoretische Physik, Universit\"at Graz, A-8010 Graz, Austria \\
              \email{elmar.biernat@uni-graz.at, wolfgang.schweiger@uni-graz.at }
           \and
           W.H. Klink \at
             Department of Physics and Astronomy, University of Iowa, Iowa City, Iowa, USA\\
\email{william-klink@uiowa.edu}
}

\date{Received: date / Accepted: date}

\maketitle

\begin{abstract}
We discuss a Poincar\'e invariant coupled-channel formalism which
is based on the point-form of relativistic quantum mechanics.
Electromagnetic scattering of an electron by a 2-body bound state
is treated as a 2-channel problem for a Bakamjian-Thomas-type mass
operator. In this way retardation effects in the photon-exchange
interaction are fully taken into account. The electromagnetic
current of the 2-body bound state is then extracted from the
one-photon-exchange optical potential. As an application we
calculate electromagnetic pion and deuteron form factors. Wrong
cluster properties, inherent in the Bakamjian-Thomas framework,
are seen to cause spurious (unphysical) contributions in the
current. These are separated and eliminated in an unambiguous way
such that one is left with a current that has all the desired
properties.

\end{abstract}

\noindent The electromagnetic structure of a few-body bound state
is usually encoded in  Lorentz invariant functions, the
electromagnetic form factors. The electromagnetic current of the
bound-state can then be written as a sum of independent Lorentz
covariants which are multiplied by these form factors. The form
factors (in the space-like region) are functions of Mandelstam $t$
($\equiv$ momentum transfer squared $q^2=-Q^2$). The theoretical
challenge is to find an appropriate expression for the bound-state
current in terms of the constituents currents. Due to binding
effects it obviously cannot be a simple sum of the constituent
currents~\cite{Siegert:1937yt}. Furthermore, the bound-state
current must transform like a four vector under Lorentz
transformations, it has to be conserved and the form factors
should be normalized correctly at $Q^2=0$.

We use the point form of relativistic  quantum mechanics which is
characterized by the property that only the four generators of
space-time translations are interaction dependent. In order to
introduce interactions into these generators in a Poincar\'e
invariant manner we make use of the, so-called, \lq\lq
Bakamjian-Thomas construction\rq\rq~\cite{BT53}. By means of this
construction the (interacting) four-momentum operator becomes a
product of an interacting mass operator times a free four-velocity
operator. It thus suffices to study just the eigenvalue problem
for the mass operator. In the point form it is most convenient to
use velocity states as basis states. Velocity states describe the
system by its overall velocity $\vec v$ and the center-of-mass
momenta $\{\vec{k}_i\}$ and canonical spin projections
$\{\mu_i\}$~\cite{Kl98} of its components (plus possible discrete
quantum numbers). Notably, the Bakamjian-Thomas approach allows
even for instantaneous interactions. Its drawback, however, is the
violation of cluster separability (macroscopic
locality)~\cite{KP91}. In the following we will focus on how this
may affect electromagnetic currents and discuss a possible
solution.

We treat elastic electron-bound-state scattering as a 2-channel
problem for a Bakam\-jian-Thomas type mass operator and use the
velocity-state representation. The emission and absorption of the
photon ($\gamma$) by the electron ($e$) and the 2 constituents
($c_1$, $c_2$) of the bound state ($C$) are described by vertex
interactions which are derived from quantum field
theory~\cite{Kl03}. Inherent in the Bakamjian-Thomas framework is
the conservation of the total four-velocity of the system which
must also be demanded at the $\gamma$-vertices. The binding force
between the constituents is described by an instantaneous
potential which is added to the operator for the free invariant
mass of the $e c_1c_2$  and the $ec_1c_2\gamma$ channels. After
elimination of the $ec_1c_2\gamma$ channel by means of a Feshbach
reduction one ends up with the 1-$\gamma$-exchange optical
potential $V_{\mathrm{opt}}$. The bound-state current
$J^\mu(\vec{k}_C^\prime,\mu_j^\prime ; \vec{k}_C,\mu_j)$ can then
be extracted from on-shell matrix elements of $V_{\mathrm{opt}}$.
$J^\mu(\vec{k}_C^\prime,\mu_j^\prime ; \vec{k}_C,\mu_j)$ is
essentially an integral over bound-state wave functions,
constituents currents and Wigner $D$ functions (see, e.g.,
Ref.~\cite{Bi09}).

For the case of a spin-0 particle like, e.g., the pion ($\pi$) it
can  be shown that $J^\mu(\vec{k}_\pi^\prime; \vec{k}_\pi)$ is
conserved and that the components of
$B_c(\vec{v})^\mu_{\,\nu}J^\nu(\vec{k}_\pi^\prime; \vec{k}_\pi)$
transform like a four-vector under Lorentz
transformations\footnote{$B_c(\vec{v})$ is a canonical
(rotationless) boost.}. It is clear that the physical pion current
should depend only on one covariant times the pion form factor
$f$, with $f$ being a function of $Q^2$. This is, however, not the
case for $J^\mu(\vec{k}_\pi^\prime; \vec{k}_\pi)$ derived in the
way we did. We rather need two covariants and 2 form factors $f$
and $b$:
 \begin{eqnarray}J^\mu(\vec{k}_\pi^\prime;
\vec{k}_\pi)=f(Q^2,s)(k_\pi^\prime+k_\pi)^\mu+b(Q^2,s)
(k_e^\prime+k_e)^\mu.\label{pioncurrent}
\end{eqnarray}
The additional (current conserving) covariant is the sum of the
electron momenta. Note that both, the physical form factor
$f(Q^2,s)$ and the spurious form factor $b(Q^2,s)$, exhibit also a
dependence on Mandelstam $s$ (or equivalently on $k=|\vec{k}_C|$).
The reason for this $s$-dependence and the spurious contribution
to the current is the violation of cluster separability in the
Bakamjian-Thomas construction. A numerical analysis shows that for
large $s$ the form factor $f(Q^2,s)$ becomes independent of $s$
and $b(Q^2,s)$ vanishes~\cite{Bi10}. In the limit $s\rightarrow
\infty$ one obtains a simple analytic expression for the pion form
factor~\cite{Bi09},
which is equivalent with the standard front form result of
Ref.~\cite{CC88}.

\begin{figure}
\includegraphics[width=6cm]{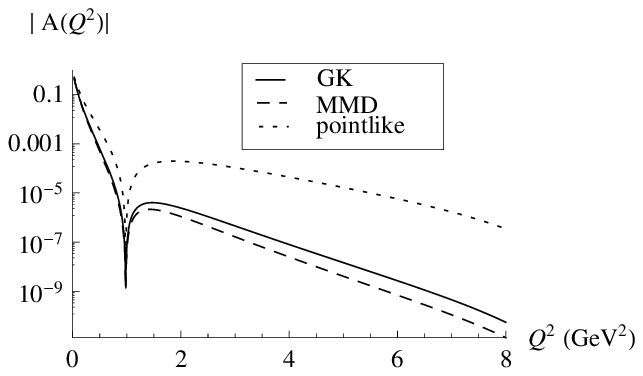}
\includegraphics[width=6cm]{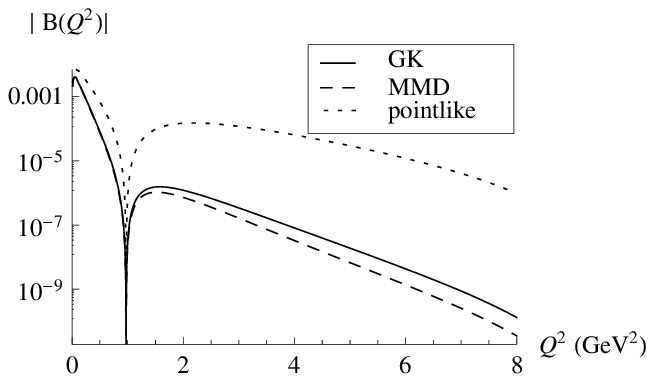}
\caption{Elastic scattering observables $A(Q^2)$ and $B(Q^2)$ for
different  parametrizations of the nucleon form factors
(GK~\cite{GK}, MMD~\cite{MMD}, point-like nucleons) for a simple
Walecka-type model. \vspace{-0.5cm}}\label{fig}
\end{figure}

For the case of a spin-1 system we consider the deuteron ($D$) as
an example. Our deuteron current
$J^\mu(\vec{k}_D^\prime,\mu_j^\prime; \vec{k}_D,\mu_j)$ can be
decomposed into 11 independent Lorentz-covariants which are
constructed from the polarization vectors
$\epsilon^{\ast\mu}(k_D',\mu_j')$ and $\epsilon^{\mu}(k_D,\mu_j)$
and the 4-momenta $(k_D+k_D')^\mu$, $(k_e+k_e')^\mu$ and $q^\mu$.
Correspondingly, we get 3 physical form factors $f_1$, $f_2$,
$g_M$ and 8 spurious form  factors. In the limit
$k\rightarrow\infty$ the 3 physical form factors become again
independent of $k$. Unlike the pion case not all spurious
contributions vanish.
This result resembles the one obtained within a covariant
light-front approach~\cite{Carbonell:1998rj}. But, most
importantly, the physical form factors can be extracted from the
current matrix elements in an unambiguous way. Using our standard
kinematics with momentum transfer in 1-direction~\cite{Bi09} we
get:
 \begin{eqnarray}
&&F_1(Q^2)= \lim_{k\rightarrow\infty} f_1(Q^2,k)=
\lim_{k\rightarrow\infty}\frac{(-1)}{2k}\left[J^0(\vec{k}_D^\prime,1;
\vec{k}_D,1)+J^0(\vec{k}_D^\prime,1;
\vec{k}_D,-1)\right], \nonumber \\
&&F_2(Q^2)=\lim_{k\rightarrow\infty}
f_2(Q^2,k)=\lim_{k\rightarrow\infty}\frac{(-2m_D^2)}{Q^2k}J^0(\vec{k}_D^\prime,1;
\vec{k}_D,-1),\\
&&G_M(Q^2)=\lim_{k\rightarrow\infty}
g_M(Q^2,k)=\lim_{k\rightarrow\infty}\frac{(-\mathrm
i)}{Q}J^2(\vec{k}_D^\prime,1; \vec{k}_D,1).\nonumber
\end{eqnarray}
The elastic scattering observables $A(Q^2)$ and $B(Q^2)$ for a
simple Walecka-type model \cite{Walecka,Bakker} are depicted in
Fig.~\ref{fig}.\footnote{This calculation of deuteron form factors
within a Walecka-type model is part of a benchmark calculation to
fix the notion of \lq\lq relativistic effects\rq\rq\  in few-body
problems.}

At the end we want to emphasize that this kind of formalism is
very general such that it is, e.g., also able to accommodate for
n-particle bound states or for exchange-current effects which are
caused by binding forces due to dynamical particle exchange.

\vspace{-0.4cm}

\begin{acknowledgements}
E.P.B. acknowledges the support of the \lq\lq Fonds zur
F\"orderung der wissenschaftlichen Forschung in \"Osterreich" (FWF
DK W1203-N16). He also thanks the province of Styria, Austria, for a PhD grant.
\end{acknowledgements}

\end{document}